\documentclass[twocolumn,amsmath,aps,superscriptaddress]{revtex4-1}

\usepackage{graphicx}
\usepackage{dcolumn}
\usepackage{multirow}
\usepackage{bm}
\usepackage{color}
\usepackage{subfigure}
\usepackage{hyperref}
\usepackage{amsmath}
\usepackage{amsfonts}

\hypersetup{colorlinks=true, citecolor=blue,pdfborder={0 0 0}}
\newcommand{\parallelsum}{\mathbin{\!/\mkern-5mu/\!}}

\makeatletter
\renewcommand{\section}{\@startsection{section}{1}{0mm}
  {-\baselineskip}{0.5\baselineskip}{\bf\center}}
\makeatother

\begin{document}
\title{The anomalous antiferromagnetic topological phase in pressurized $\mathrm{SmB_6}$}

\author{Kai-Wei Chang}
\affiliation{Department of Physics, National Taiwan University, Taipei 10617, Taiwan}
\affiliation{Nano Science and Technology Program, Taiwan International Graduate Program, Academia Sinica, Taipei 11529, Taiwan and National Taiwan University, Taipei 10617, Taiwan}
\affiliation{Institute of Physics, Academia Sinica, Taipei 11529, Taiwan}
\author{Peng-Jen Chen}
\email{pjchen1015@gate.sinica.edu.tw}
\affiliation{Institute of Physics, Academia Sinica, Taipei 11529, Taiwan}

\begin{abstract}
  Antiferromagnetic materials, whose time-reversal symmetry is broken, can be classified into the $Z_2$ topology if they respect some specific symmetry. Since the theoretical proposal, however, no materials have been found to host the antiferromagnetic topological (AFT) phase to date. Here, for the first time, we demonstrate that the topological Kondo insulator $\mathrm{SmB_6}$ can be an AFT system when pressurized to undergo an antiferromagnetic phase transition. In addition to propose the possible candidate for an AFT material, in this work we also illustrate the anomalous topological surface states of the AFT phase which has not been discussed before. Originating from the interplay between the topological properties and the antiferromagnetic surface magnetization, the topological surface states of the AFT phase behave differently as compared with those of a topological insulator. Besides, the AFT insulators are also found promising in the generation of tunable spin currents, which is an important application in spintronics.
\end{abstract}

\maketitle
The topological phase of a material is an exotic quantum state that can host interesting properties at boundaries. For instance, a topological insulator (TI), characterized by the $Z_2$ invariant, exhibits gapless surface or edge states protected by the time-reversal symmetry (TRS). These topologically protected surface states are robust against small perturbations as long as the TRS is preserved. The $Z_2$ invariant is defined for a set of bands that is quarantined by a continuous gap, a region in momentum space that separates this set of bands from all others. That is to say, metallic systems having a continuous gap can also have defined $Z_2$ invariant \cite{Hsie,Saka,Guan,Chen}, although insulators that possess a real energy gap are more common.

In general, the $Z_2$ invariant is known to be undefined in a TRS-broken system as magnetism sets in. Nonetheless, Mong {\it et al.} \cite{Mong} proposed that some specific antiferromagnetic systems, in which the TRS is broken while the combined symmetry $\mathcal{S}$ $=\Theta T_{1/2}$ is preserved, can also be classified into the $Z_2$ topological state. ($\Theta$ is the TRS operator, and $T_{1/2}$ is the translation operation along a crystal axis by half of the corresponding lattice constant. In this work, the translation is chosen to be along $\hat{x}$.) Different from the typical three-dimensional TIs, the $Z_2$ invariant of this antiferromagnetic topological (AFT) phase is defined only on the $k_x$ = 0 plane (in our case), where $S^2 = -1$ resembling $\Theta^2 = -1$ in the TIs. In other words, an AFT system is a three-dimensional material that possesses a two-dimensional $Z_2$ invariant. Also, the gapless TSSs only exist on the $\mathcal{S}$-invariant surfaces where the $Z_2$ invariant is defined. However, since the proposal of the theory, no materials have been found to reveal the AFT phase, which in turn prohibits detailed investigation of a real AFT system.

A previous work suggested that the rare earth compound, GdBiPt, could be a possible candidate for an AFT system \cite{Mull}. Although the antiferromagnetic state has been experimentally confirmed, the topological property of GdBiPt is not studied. This puts the suggested AFT phase in GdBiPt in an elusive position. Moreover, GdBiPt is a gapless semimetal, disobeying the requirement that a continuous gap must exist for a defined $Z_2$ invariant. As a result, the AFT phase in GdBiPt may need more careful examination. On the other hand, $\mathrm{SmB_6}$ has attracted much attention among the rare earth compounds for being a topological Kondo insulator \cite{Dzer,Lu,Neup,Jian,Kim,Xu} and for some observed phenomena that have not been fully settled, including the low-temperature conductivity \cite{Ment,Nick}, quantum oscillations \cite{Li,Tan,Erte}, Kondo breakdown \cite{Erte,Alex,Vale}, and so on.

In this work, we demonstrate that the pressurized $\mathrm{SmB_6}$ would serve as a better candidate for an AFT system. At ambient condition, the nonmagnetic $\mathrm{SmB_6}$ has been confirmed to exhibit nontrivial $Z_2$ topology \cite{Dzer,Lu,Neup,Jian,Kim,Xu}. The hybridization between the localized $4f$-orbitals and the itinerant $5d$-orbitals of Sm opens up a gap, making it a topological Kondo insulator. Furthermore, $\mathrm{SmB_6}$ turns into the magnetic state upon pressurization (at $P \geq 6$ GPa) \cite{Barl,Derr,Nish,Butc,Zhou}. Although experimentally it is not clear whether it is ferromagnetic or antiferromagnetic, our first-principles calculations reveal that the A-type antiferromagnetic (A-AFM) configuration is the ground state of the pressurized $\mathrm{SmB_6}$. This result is partly supported by Si {\it et. al.} \cite{Si} that the developed magnetic order is likely to be antiferromagnetic because of the observed (nearly) second-order phase transition. Our first-principles calculations further reveal that $\mathrm{SmB_6}$ possesses the following important ingredients for being an AFT system. First, the antiferromagnetic state shows a continuous gap, which is crucial for the $Z_2$ invariant to be defined. Second, the band inversion is still present. Indeed, the parity analysis at $k_x$ = 0 clearly indicates the nontrivial $Z_2$ topology of this plane. The TSSs are found to exist on the $\mathcal{S}$-invariant surfaces as required. Therefore, the pressurized antiferromagnetic $\mathrm{SmB_6}$ exhibits AFT phase and may serve as a good model system to study the interesting properties that an AFT system reveals.

\begin{figure}[t]
\includegraphics [width=8.5cm] {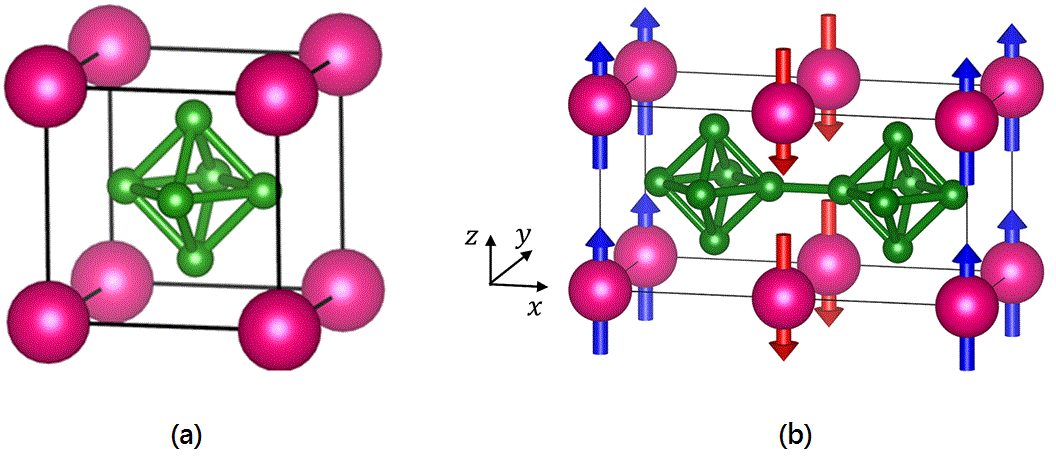}
\caption{\small{The crystal structure of $\mathrm{SmB_6}$. (a) Nonmagnetic (cubic) and (b) A-AFM (orthorhombic) $\mathrm{SmB_6}$. The smaller green spheres represent the B atoms and the larger pink spheres represent the Sm atom with arrows indicating the spins. In this work we choose the $\hat{x}$-direction as the doubled axis for the A-AFM configuration.}}
\label{crystal}
\end{figure}

\begin{table}[b]
\begin{center}
\caption{\small{The parities of $\mathrm{SmB_6}$ at the TRIM. The parities of the nonmagnetic (at 0 GPa) and the A-AFM $\mathrm{SmB_6}$ (at 8 GPa) at the TRIM. For the A-AFM phase, only the parities at TRIM with $k_x = 0$, the $\mathcal{S}$-invariant momenta, are shown.}}
\label{table:parity}
\begin{ruledtabular}
\begin{tabular}{cccccccccc}
\multicolumn{1}{c}{} &
\multicolumn{4}{c}{0 GPa} &
\multicolumn{1}{c}{} &
\multicolumn{4}{c}{8 GPa}\\ \cline{2-10}
                 & $\Gamma$ & {\it 3X} & {\it 3M} & {\it R} &  & $\Gamma$ & {\it Y} & {\it Z} & {\it T} \\ \hline
{\it U} = 0.0 eV &     +    &    $-$   &     +    &    +    &  &    $-$   &   $-$   &   $-$   &    +    \\ \hline
{\it U} = 4.0 eV &     +    &    $-$   &     +    &    +    &  &    $-$   &   $-$   &   $-$   &    +    \\ \hline
{\it U} = 8.0 eV &     +    &    $-$   &     +    &    +    &  &    $-$   &   $-$   &   $-$   &    +    \\ 
\end{tabular}
\end{ruledtabular}
\end{center}
\end{table}

\section{RESULTS}
\subsection{Crystal and electronic structures}
\begin{figure}[t]
\includegraphics [width=9cm] {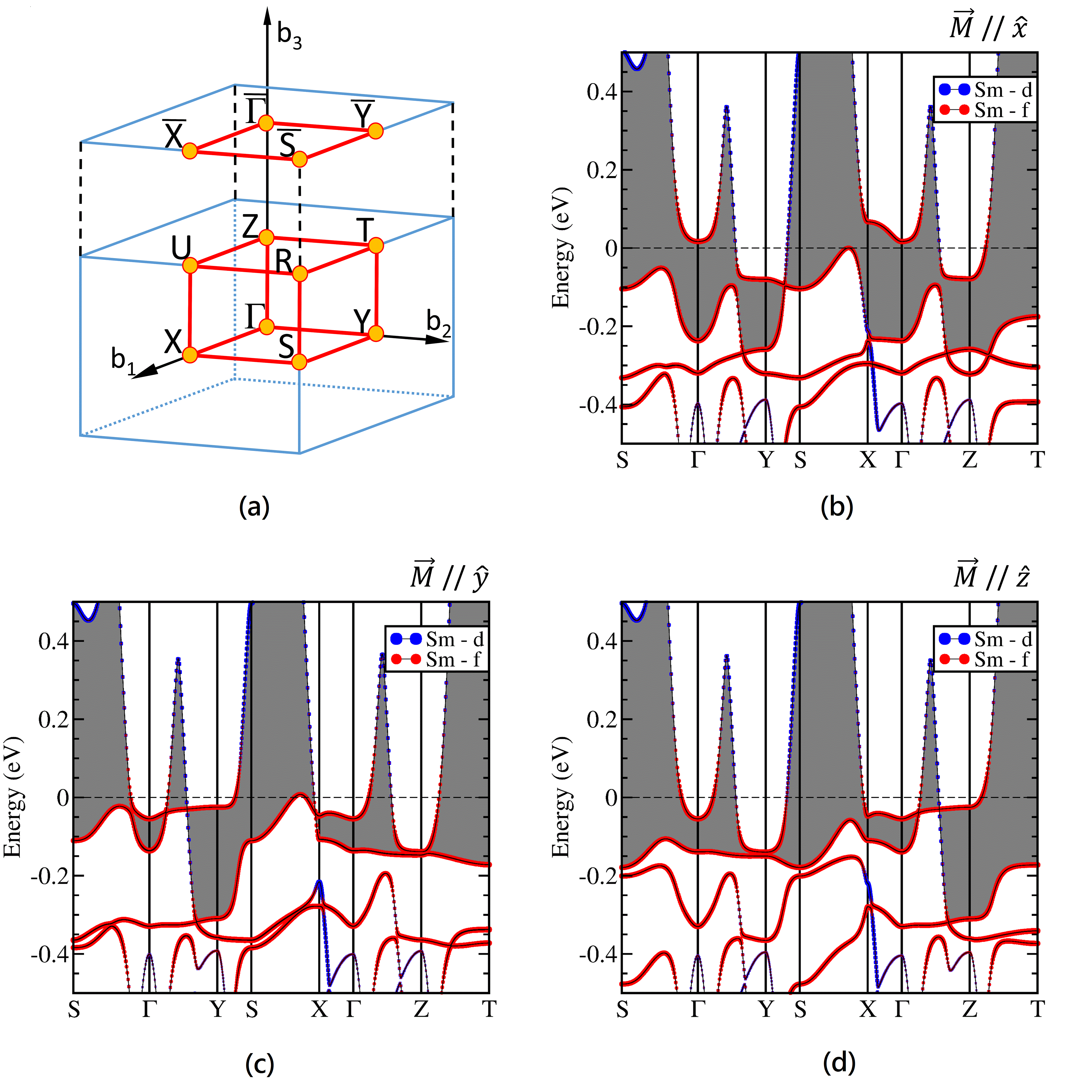}
\caption{\small{The bulk band structure of A-AFM $\mathrm{SmB_6}$. (a) The schematic illustration of the bulk and projected surface Brillouin zone of an orthorhombic crystal. (b)-(d) The bulk band structures of the A-AFM $\mathrm{SmB_6}$ with (b) $\vec{M} \parallelsum \hat{x}$ (c) $\vec{M} \parallelsum \hat{y}$, and (d) $\vec{M} \parallelsum \hat{z}$. Although tiny in some regions, the continuous gaps (the painted regions) do exist in (b)-(d).}}
\label{band-bulk}
\end{figure}

At ambient condition, $\mathrm{SmB_6}$ crystallizes in a cubic phase as shown in Fig. \ref{crystal}(a) without forming any magnetic order. For systems with inversion symmetry, the $Z_2$ invariant can be deduced from the knowledge of parities at the time-reversal invariant momenta (TRIM) \cite{Fu}. The computed parities of $\mathrm{SmB_6}$ at the TRIM are listed in Table \ref{table:parity}. Consistent with the previous works \cite{Dzer,Lu,Neup,Jian,Kim,Xu}, $\mathrm{SmB_6}$ is a strong TI at ambient condition. Since $\mathrm{SmB_6}$ is considered strongly correlated, different values of the on-site Coulomb {\it U} are used to study the topological phase. It turns out that the topological phase is independent of the values of {\it U}, in agreement with the results reported in Refs. \cite{Lu,Kang,Chan}. As the pressure increases, it undergoes a magnetic phase transition at the critical pressure of 6 $\sim$ 10 GPa \cite{Barl,Derr,Nish,Butc,Zhou}. As mentioned previously, while the actual magnetic ordering is not clarified experimentally, our first-principles calculations indicate that the A-AFM [with $\vec{M} \parallelsum \hat{x}$ as shown in Fig. \ref{crystal}(b)] is the ground state of the pressurized $\mathrm{SmB_6}$, supported by the inference made in Ref. \cite{Si}. Figure \ref{band-bulk} (b)-(d) displays the bulk band structure of the A-AFM $\mathrm{SmB_6}$. Shown in Fig. \ref{band-bulk}(b) is the band structure when $\vec{M} \parallelsum \hat{x}$. The other two cases with $\vec{M} \parallelsum \hat{y}$ and $\vec{M} \parallelsum \hat{z}$ constitute equivalent bulk systems owing to the symmetry. Their band structures are shown in Fig. \ref{band-bulk}(c)(d) where identical results are seen when $k_y$ and $k_z$ are interchanged. Important to mention, continuous gaps (the painted regions) due to the interactions of Sm-$d$ and -$f$ bands are present in all cases, making the $Z_2$ invariant defined in all A-AFM configurations.

For the TRS-protected systems with inversion symmetry, the $Z_2$ invariant can be deduced from the knowledge of parities at the TRIM. For an AFT system whose protecting symmetry is $S$ rather than the TRS, this argument still applies since $S$ is isomorphic to $\Theta$ at $k_x=0$ (in our case). The parities at the TRIM of the A-AFM $\mathrm{SmB_6}$ are computed and shown in Table \ref{table:parity}. Apparently, the parities at the four TRIM on the $k_x$ = 0 plane yield a nontrivial $Z_2$, indicating the AFT phase. It is also found that the AFT phase of $\mathrm{SmB_6}$ is qualitatively independent of the pressure, so here we choose $P = 8$ GPa to demonstrate the results. (The results at $P = 6$ and 10 GPa are shown in Supplementary Figs. 1 and 2 for comparison.) Again, the parities, as well as the $Z_2$ invariant, are shown to be independent of the effect of correlation as listed in Table \ref{table:parity}. As a result, we use $U$ = 0 in the following calculations of $\mathrm{SmB_6}$ to demonstrate the topological properties of an AFT system.

\begin{figure}[t]
\includegraphics [width=8.5cm] {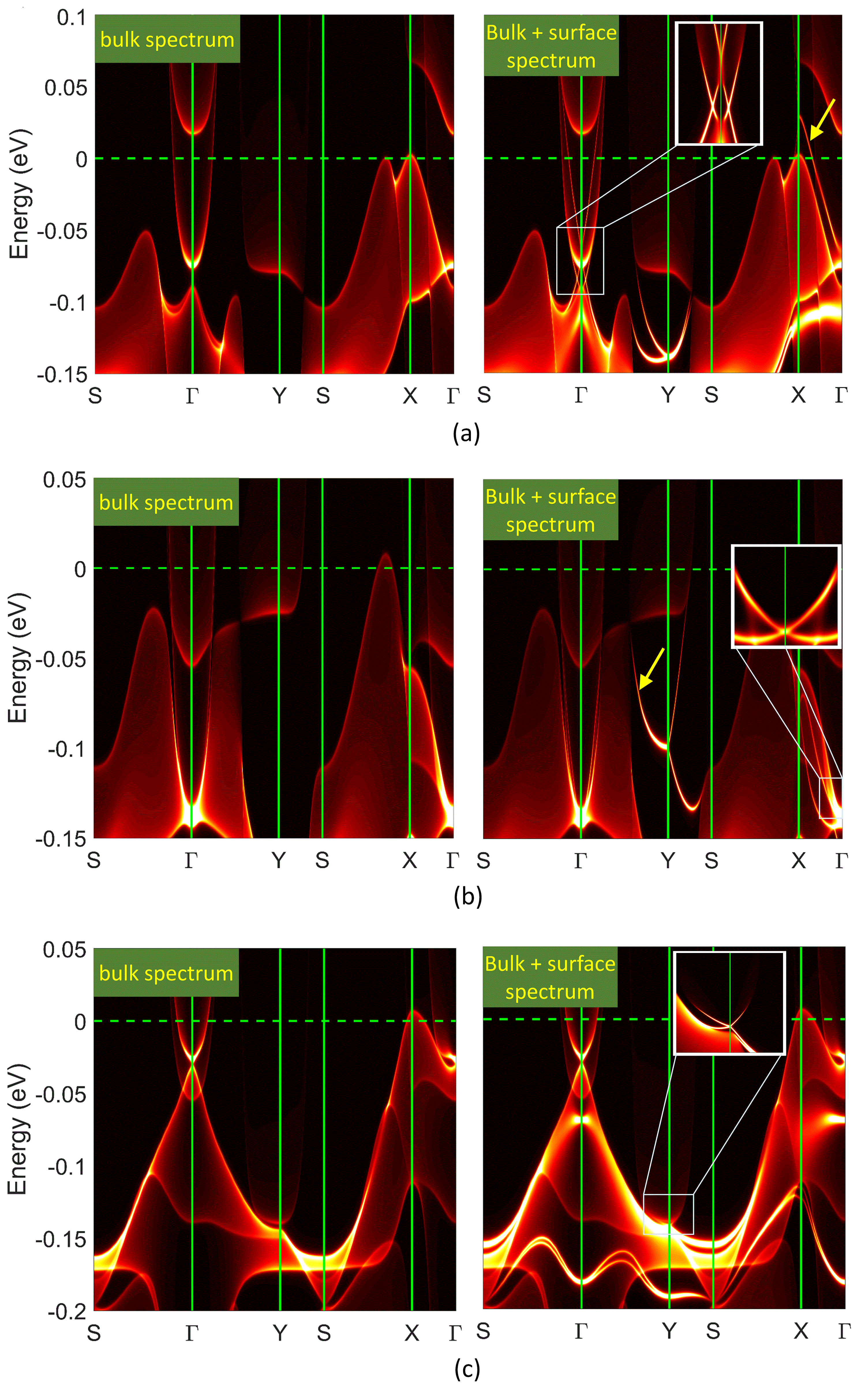}
\caption{\small{The spectra of the semi-infinite A-AFM $\mathrm{SmB_6}$ slab. (a) $\vec{M} \parallelsum \hat{x}$, (b) $\vec{M} \parallelsum \hat{y}$, and (c) $\vec{M} \parallelsum \hat{z}$. Left and right panels display the spectral weights contributed from the bulk only and bulk plus surface, respectively. The insets are the zoomed-in spectra in the selected regions. Only the surface spectral weights are shown in all insets to more clearly depict the TSSs. The yellow arrows indicate the doubly degenerate TSSs.}}
\label{band-surf}
\end{figure}

\subsection{The topological surface states}

As shown in Fig. \ref{band-surf}, the surface band structures clearly reveal the presence of the TSSs, regardless of the direction of $\vec{M}$. For the case of in-plane magnetization shown in Fig. \ref{band-surf}(a)(b), three TSSs are present: two of them cross at $\overline{\Gamma}$ and the remaining one crosses at $\overline{Y}$. (The overline denotes the high symmetry points in the surface Brillouin zone.) Compared with the bulk band structure shown in Fig. \ref{band-bulk}(b), the two TSSs crossing at $\overline{\Gamma}$ come from the bulk bands at $\Gamma$ and {\it Z}, respectively. For $\vec{M} \parallelsum \hat{y}$, the one coming from {\it Z} is merged into the bulk bands so it can be hardly seen in the band structure. The out-of-plane magnetization, on the other hand, shows only one TSS crossing at $\overline{Y}$. Together with the calculated $Z_2$ invariant listed in Table \ref{table:parity}, the AFT phase of the antiferromagnetic $\mathrm{SmB_6}$ is verified, regardless of the direction of $\vec{M}$. An interesting thing to note is that the TSSs have two forms when $\vec{M}$ is in-plane: the arc-like open orbits and closed loops. For the arc-like ones, the TSSs are doubly degenerate (marked by the yellow arrows in the figure), rather than being singly degenerate as in the TIs, when the two arcs meet. The degeneracy takes place at $k_y = 0$ ($k_x = 0$) when $\vec{M} \parallelsum \hat{x}$ ($\vec{M} \parallelsum \hat{y}$). For the out-of-plane case, the TSSs are singly degenerate in all directions, which is consistent with the result reported in Ref. \cite{Mong}. The double degeneracy of TSSs is an interesting behavior that has not been reported in the $Z_2$-nontrivial systems before and will be discussed later.

\begin{figure*}[t]
\includegraphics [width=15.0cm] {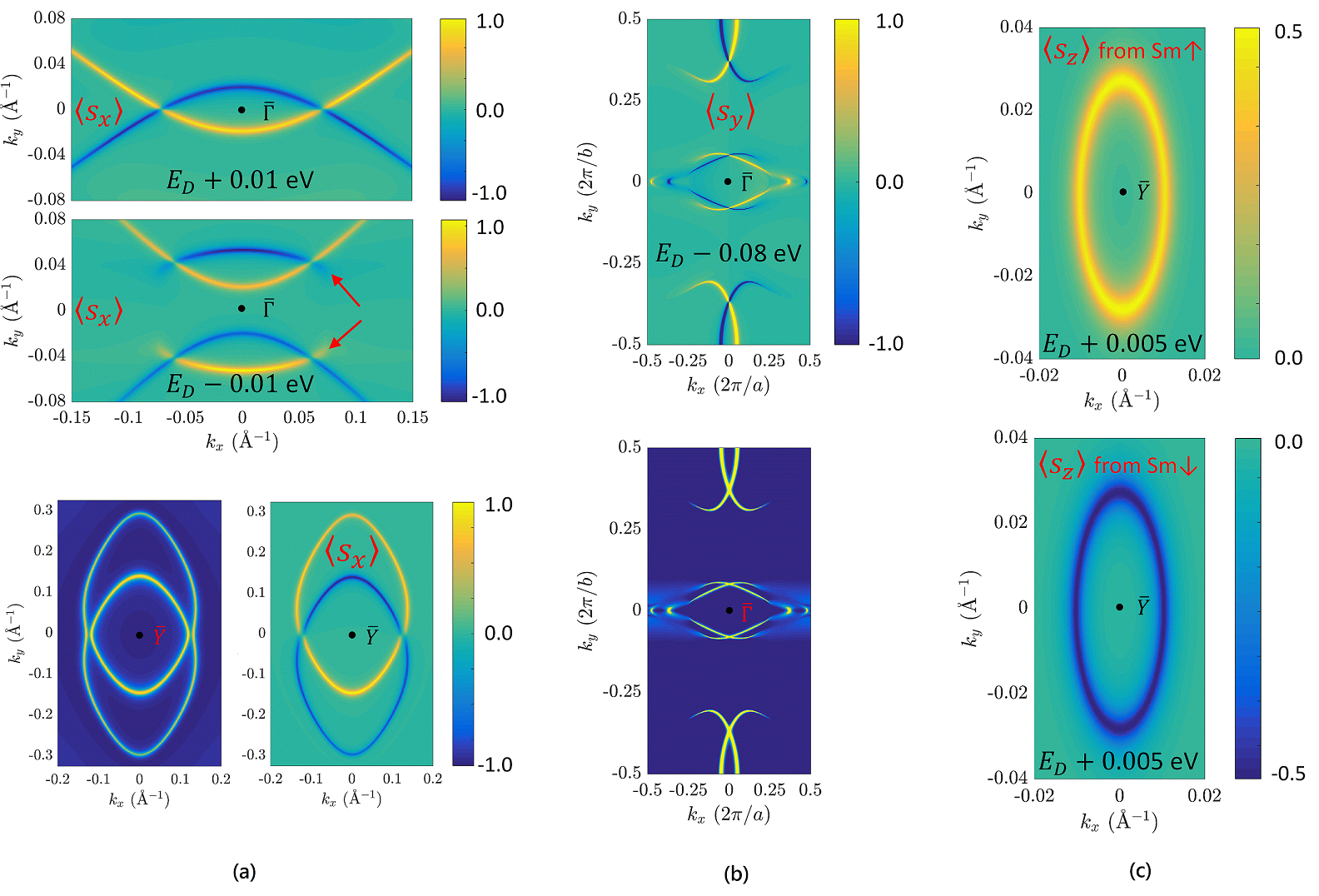}
\caption{\small{The constant energy contours of the TSSs. (a) $\vec{M} \parallelsum \hat{x}$. The open (arc-like) TSSs around $\overline{\Gamma}$ and the closed TSSs around $\overline{Y}$ are shown in the upper and lower panels, respectively. The left part of the lower panel represents the spectrum, from which the closed loops of TSSs can be found. The spin states of the closed TSSs around $\overline{Y}$ change sign when crossing $k_y = \pi$/b. Note that in the lower panel the ``zero'' of the $k_y$-axis is chosen to be $k_y = \pi$/b (zone boundary). Red arrows in the upper panel indicate the other TSS around $\Gamma$ with lower $E_D$. (b) $\vec{M} \parallelsum$ $\hat{y}$. The open (arc-like) and closed TSSs are present around $\overline{Y}$ and $\overline{\Gamma}$, respectively, as the upper panel shows. The lower panel represents the spectrum. The spin states of the closed TSSs around $\overline{\Gamma}$ change sign when crossing $k_x = 0$. For (a) and (b), the TSSs are fully spin-polarized in the direction along $\vec{M}$. (c) $\vec{M} \parallelsum$ $\hat{z}$. The upper and lower panels represent the partial $\langle s_z \rangle$ contributed from the two Sm atoms with opposite magnetization. The spins with equal magnitude but opposite sign result in zero $\langle s_z \rangle$ in total. The ``zero'' of the $k_y$-axis is, again, chosen to be $k_y = \pi$/b (zone boundary).}}
\label{spin-texture}
\end{figure*}

The spin texture of the TSSs in the AFT phase is another interesting issue to investigate. For typical TIs, these TSSs show spin helicity. However, for an AFT system where the surfaces hosting the TSSs remain antiferromagnetically ordered, the coupling between the TSSs and the local moment of the surface atoms is not straightforward to understand. As shown in Fig. \ref{spin-texture}(a)(b) for the case of in-plane magnetization, the spins of the TSSs in the AFT phase align parallel to the direction of $\vec{M}$, rather than perpendicular to $\vec{k}$. This can be understood by the antiferromagnetic arrangement on the surface that forces all spins to align along $\vec{M}$ to gain the Zeeman energy. The development of the other in-plane component is not favored because of the much higher energy cost of the Zeeman energy if the spins are rotating in the $(k_x,k_y)$ plane, which in turn forbids the formation of helical spins. For the out-of-plane magnetization shown in Fig. \ref{spin-texture}(c), the TSSs form an usual Dirac cone as those of a TI do. Interestingly, however, the TSSs are non-spin-polarized due to the equal contributions from the two magnetic sublattices with up and down spins. Under the influence of the surface magnetization, the spins of each magnetic sublattice also align along $\vec{M}$. Shown in Fig. \ref{spin-texture}(c) is the upper part of the TSSs that has energy higher than that of the Dirac point ($E_D$). For the lower part with $E < E_D$, the behaviors are similar only that the spins of each magnetic sublattice are reversed, preserving zero $\langle s_z \rangle$. To our knowledge, this is the first kind of TSSs that exhibit zero spin polarization.

\section{DISCUSSION}

\begin{figure}[t]
\includegraphics [width=8cm] {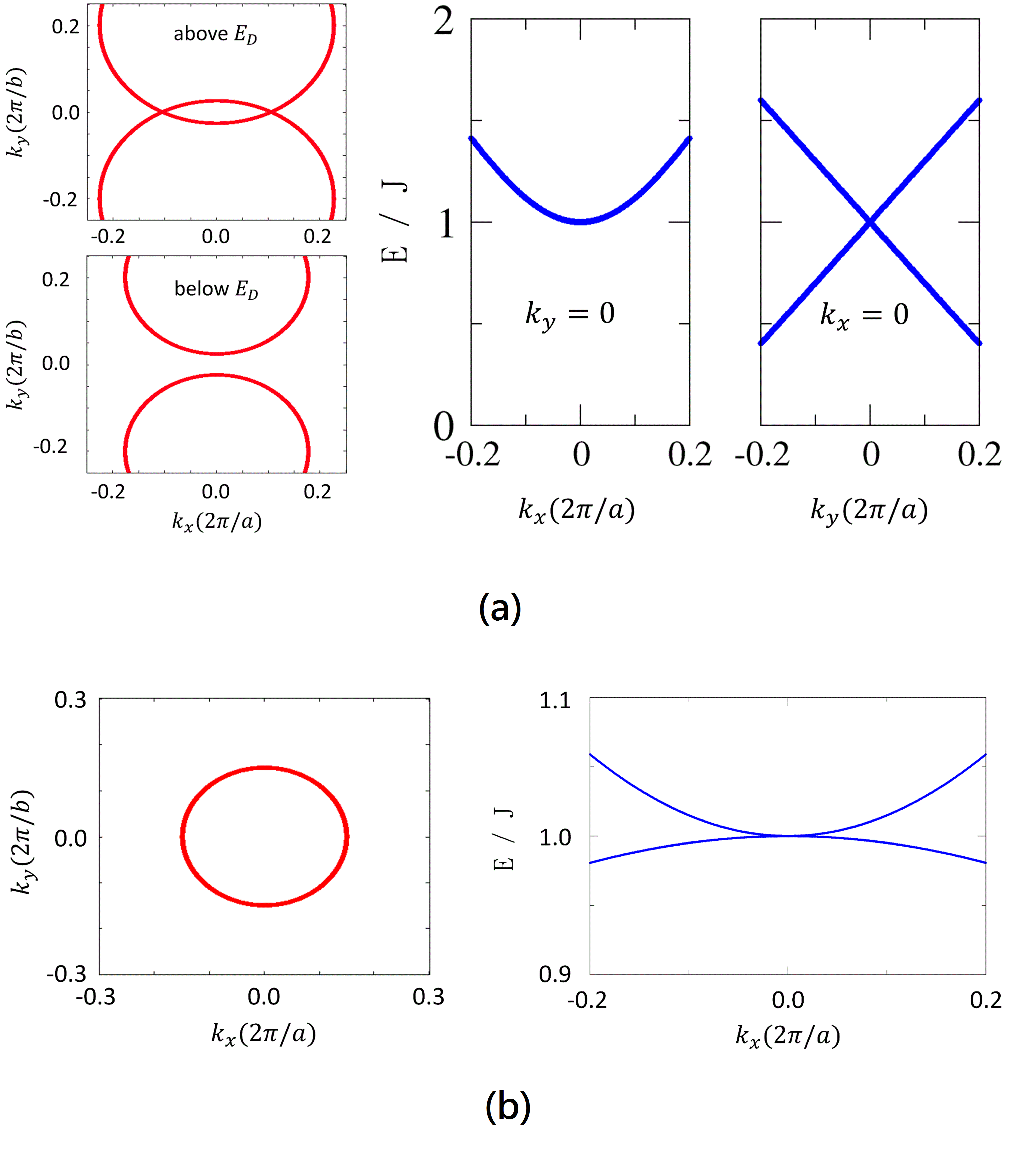}
\caption{\small{Energy bands calculated from the 4$\times$4 phenomenological model. The constant energy contours (left) and the band structures (right) as (a) $\vec{M} \parallelsum \hat{x}$ with $J/\lambda$ = 1.0/3.0  and (b) $\vec{M} \parallelsum \hat{z}$ with $J/\lambda$ = 1.0/1.0. The upper and lower panels of the left part of (a) represent the energies above and below $E_D$, respectively.}}
\label{model-band}
\end{figure}

It is natural to look at the Rashba effect on an antiferromagnetic surface when discussing the TSSs of an AFT system. For this reason, we construct a phenomenological 4$\times$4 Hamiltonian as below.
\begin{eqnarray*}
H(\vec{k}) & = & [\lambda(\vec{\mathit{k}}\times\hat{z})\cdot\sigma + |\vec{k}_{\perp}|^2\sigma_{\vec{M}}]\otimes\mathbb{I}_{2\times2} + {J \vec{m}\cdot \sigma\otimes\tau_z},
\end{eqnarray*}
where $\lambda$ describes the strength of spin-orbit coupling (SOC) , $\sigma$ ($\tau$) the spin (orbital) Pauli matrices (the subscript $\vec{M}$ denotes the one parallel to it), $\vec{k}_{\perp}$ the momentum perpendicular to $\vec{M}$, $J$ the Zeeman coupling strength, and $\vec{m}$ a unit vector along $\vec{M}$. The first, second and third terms represent the Rashba, kinetic, and Zeeman terms, respectively. The effect of $J$ is to introduce the Zeeman splitting of bands. Without loss of generality, only the bands with positive energies are shown in Fig. \ref{model-band}. For the case of $\vec{M} \parallelsum \hat{x}$ shown in Fig. \ref{model-band}(a), it shifts the bands along $k_y$ through the coupling to the Rashba term, $k_y\sigma_x$. Crossings of the two TSSs, and hence the doubly degenerate TSSs along $\overline{\Gamma}$-$\overline{X}$, occur as $E > E_D$. Apparently, the shifts in $k_y$ and the double degeneracy of the TSSs along $\overline{\Gamma}$-$\overline{X} $as $\vec{M} \parallelsum \hat{x}$ originate from the antiferromagnetic magnetization of the surface atoms; the antiferromagnetic surface atoms give rise to the Zeeman splitting that couples to the in-plane momenta. This is an interesting phenomenon that has not been discussed and reported before in other kinds of topological systems. The TSSs when $\vec{M} \parallelsum \hat{y}$ can be explained in a similar way. In this case, the coupling to $k_x\sigma_y$ leads to the shift in $k_x$ direction. The band dispersion as $\vec{M} \parallelsum \hat{z}$ is also reproduced as shown in Fig. \ref{model-band}(b). It is worth mentioning that the separated bands shown in the right panel of Fig. \ref{model-band}(b) is due to their opposite signs of $\sigma_{\vec{M}}$ through the kinetic term. However, despite the success in describing the band dispersion, the above-mentioned simple phenomenological model fails to give correct spins that are forced to align along $\vec{M}$ when it is in-plane; when $\vec{M} \parallelsum \hat{x}$, for example, the spin-momentum locking due to the Rashba term induces the tendency of helical spins that gives rise to smaller but comparable $\langle s_y \rangle$, which is inconsistent with what we have from the first-principles calculations. Thus, a new model beyond the phenomenological base, e.g. with the inclusion of the nontrivial electronic structure, is required to further investigate the intricate problems on a surface showing AFT phase.

Besides the proposal of a possible candidate for an AFT system, our work is dedicated to investigating the topological properties of the AFT phase. Based on our results, the spin polarization of the TSSs in the case of in-plane magnetization naturally generates a spin current if AFT {\it insulators} are found. Also, by shifting the chemical potential above $E_D$, the spin current can be tuned down due to the crossed TSSs with opposite spins; for a given direction of a spin current, more and more electrons with opposite spin will be moving along this direction as the chemical potential is increasing. This tunable spin current generated by the robust TSSs opens up a new route for the application in spintronics. Furthermore, a spin current flowing in a material with strong SOC, which is true in most topological materials, is known to induce a charge current through the inverse spin Hall effect (ISHE). In general, the direction of the ISHE-induced charge flow is perpendicular to both the spin polarization and the momentum of the spin current because of the Rashba interactions. As previously discussed, the Rashba interactions seem to play a secondary role on the surface of an AFT system. Under this circumstance, the transport behaviors of the ISHE in an AFT insulator may call for further investigation. After all the theoretical discussions of the AFT phase, the following question is the experimental observation. Based upon the above discussions, we suggest that the generation of a spin current should provide a possible way of detecting the TSSs of an AFT system. The existence of the TSSs could be evidenced through the detection of a spin current with polarization parallel to the magnetization, provided that the TSSs contribute to the transport.

In addition to the underlying mechanism, there are also some interesting fields that can be linked with the anomalous properties of an AFT system. For example, low-energy excitation in spins is an important issue in both the theoretical interests and the application in spintronics. For an AFT system with strong SOC and magnetization on surfaces where the inversion symmetry is broken, Dyzaloshinskii-Moriya interactions can possibly take place. Some excitations of collective modes, e.g. spiral spin waves and the Skyrmions, are allowed to happen. How the TSSs affect or couple to these excitations would be an interesting issue to study. Another example would be the Landau level splitting. The Landau level splitting, as well as the associated magnetotransport properties, may show anomalous behavior as the AFT properties come into play, which may be different from the reported results of the magnetically confined two-dimensional Dirac fermions in a bulk antiferromagnet $\mathrm{EuMnBi_2}$ \cite{Masu}.

Before closing, we would like to mention that the strong correlation, which is not considered in this work, may affect the band structure. This can possibly change the ranges of energy where the TSSs are present. However, the anomalous features of the TSSs of an AFT system that our work indicates should still apply. Through our preliminary work, we hope to invoke more efforts to disclose the underlying mechanism for the intricate topological behaviors of the AFT phase. The thorough understanding of the AFT phase can enable us to further study the interplay between magnetism and topology, such as the low-energy spin excitations and spin transport.

In summary, our work opens up a new research field in the magnetic topological materials. Using the pressurized A-AFM $\mathrm{SmB_6}$ as an example, we demonstrate the interesting yet unknown behaviors of the TSSs of an AFT system. For the case of in-plane magnetization, the TSSs can form arc-like open orbits and double degeneracy occurs when two arcs meet each other. The spins of the TSSs are forced to align along the direction of $\vec{M}$, losing the characteristic of helical spins. For the out-of-plane magnetization, the spin helicity of the TSSs are present because the slightly tilted spins gain the Rashba energy at the cost of low Zeeman energy. Our work not only points out the possible realization of the AFT phase in the pressurized $\mathrm{SmB_6}$, verifying the theoretically proposed topological phase, but, more importantly, unravels the exotic topological behaviors of the TSSs of an AFT system. As suggested, AFT insulators can be used to generate the tunable spin current, which can have important application in spintronics.

\section{Methods}
The first-principles calculations are carried out using QUANTUM ESPRESSO (QE) code \cite{QUAN} with norm-conserving PBE functionals generated by atomic code in QE distribution. The Sm-$5p$ electrons are included as the semicore valence sates. The energy cut for the plane wave expansion is 100 Ry to deal with the hard pseudopotential of Sm. Experimental lattice constants of the pressurized $\mathrm{SmB_6}$ \cite{Nish} are adopted in all calculations with SOC taken into account. The surface band structures are calculated using the semi-infinite slab model based on the Green's functions \cite{San1,San2}. Wannier functions obtained from the Wannier90 code \cite{Most} are used to compute the inter- and intra- layer couplings of the slab.

{}

\section{Acknowledgements}
P.J.C. thanks the insightful discussions with Ting-Kuo Lee. The authors acknowledge the National Center for High-Performance Computing (NCHC). This work is supported by the Ministry of Science and Technology (MOST) and Academia Sinca.

\section{Author contributions}
P.J.C. conceived and supervised the project and wrote the paper. K.W.C. and P.J.C. did the calculations and analyzed the data. K.W.C. and P.J.C. contributed equally to this work.

\end {document}